\documentclass[pamm,a4paper,fleqn,oneside]{w-art}
\usepackage{times}
\usepackage{w-thm}
\theoremstyle{plain}

\theoremstyle{definition}

\usepackage[]{graphicx}
\chardef\bslash=`\\ % p. 424, TeXbook

\newcommand{\ud}{\mathrm{d}}

\begin{document}
%%    The information for the title page will be placed between
%%    \begin{document} and \maketitle. The order of most entries
%%    is determined by the class file and can not be changed by
%%    rearranging them. The maketitle command follows after the
%%    abstract.
%%
%%    The following commands will be updated by the publisher:
%%
%%    \renewcommand{\copyrightyear}{2006}
%%    \DOIsuffix{pamm.20061zzzz}
%%    \Volume{6} \Year{2006} \pagespan{xxx}{yyy}
%%

%%%%%%%%%%%%%%%%%%%%%%%%%%%%%%%%%%%%%%%%%%%%%%%%%%%%%%%%%%%%%%%%%%%%%%%%%%
\begin{minipage}[t]{180mm}
\thispagestyle{empty}
\vspace{20mm}

\begin{center}
{\Large\bf Fusion modeling in plasma physics: Vlasov-like systems }

\vspace{20mm}

{\large\bf Antonina N. Fedorova, Michael G. Zeitlin}

\vspace{20mm}

Mathematical Methods in Mechanics Group \\

Institute of Problems of Mechanical Engineering (IPME RAS)\\ 

Russian Academy of Sciences \\

Russia, 199178, St. Petersburg, V.O., Bolshoj pr., 61\\

zeitlin@math.ipme.ru, anton@math.ipme.ru\\
         
http://www.ipme.ru/zeitlin.html\\

http://mp.ipme.ru/zeitlin.html

\vspace{20mm}
{\bf Abstract}

\vspace{10mm}

\begin{tabular}{p{100mm}}

  The methods developed by authors are applied to some 
reductions of BBGKY hierarchy, namely, various examples of Vlasov-like systems
which are important both for fusion modeling and for particular physical problem
s 
related to plasma/beam physics. We mostly concentrate on phenomena of 
localization and pattern formation. 

\vspace{10mm}

Presented: GAMM Meeting, 2006, Berlin, Germany.
\vspace{5mm}

Published: Proc. Appl. Math. Mech. (PAMM), {\bf 6}, 625-626, Wiley-VCH, 2006,\\

\vspace{5mm}

 {\bf DOI} 10.1002/pamm.200610293

\end{tabular}

\end{center}
\end{minipage}
\newpage

\title{Fusion modeling in plasma physics: Vlasov-like systems}

%% Please delete not needed author entries.
%% If there is only one address the marker \inst{x} may be omitted.
%% Information for the first author.
%\author{First Author\footnote{Corresponding
%     author: e-mail: {\sf x.y@xxx.yyy.zz}, Phone: +00\,999\,999\,999,
%     Fax: +00\,999\,999\,999}\inst{1}} \address[\inst{1}]{First address}
%%
%%    Information for the second author
%\author{Second Author\footnote{Second author footnote.}\inst{1,2}}
%\address[\inst{2}]{Second address}
%%
%%    Information for the third author
%\author{Third Author\footnote{Third author footnote.}\inst{2}}
%%
%%    \dedicatory{This is a dedicatory.}

\author{Antonina N. Fedorova\footnote{Corresponding
     author: e-mail: {\sf anton@math.ipme.ru}, http://www.ipme.ru/zeitlin.html,
  http://mp.ipme.ru/zeitlin.html}}
\author{Michael G. Zeitlin}

\address[]{IPME RAS, St.~Petersburg, V.O. Bolshoj pr., 61, 199178, Russia}

\begin{abstract}      
   The methods developed by authors are applied to some 
reductions of BBGKY hierarchy, namely, various examples of Vlasov-like systems
which are important both for fusion modeling and for particular physical problems 
related to plasma/beam physics. We mostly concentrate on phenomena of 
localization and pattern formation. 
\end{abstract}

%% maketitle must follow the abstract.
\maketitle                   % Produces the title.

         We sketch the applications of our approach based on variational 
multiresolution technique [1], [2] to the systems with collective type 
behaviour described by some forms of Vlasov-Poisson/Maxwell equations, some important 
reduction of general BBGKY hierarchy [3]. 
Such an approach may be useful in all models in which it is possible and reasonable to 
reduce all complicated problems related to statistical distributions to the problems described 
by the systems of nonlinear ordinary/partial differential/integral equations with or without 
some (functional) constraints. In periodic accelerators and transport systems at the high beam 
currents and charge densities the effects of the intense self-fields, which are produced by the 
beam space charge and currents, determinine (possible) equilibrium states, stability and 
transport properties according to underlying nonlinear 
dynamics. The dynamics of such space-charge dominated 
high brightness beam systems can provide the understanding of the 
instability phenomena such as emittance growth, mismatch, halo formation related to the
complicated behaviour of underlying hidden nonlinear modes outside of perturbative
KAM regions [3].
Our analysis based on the variational-wavelet approach allows to consider 
polynomial and rational type of nonlinearities [1], [2]. 
In some sense in this particular case this approach is direct generalization of 
traditional nonlinear $\delta F$  approach [3] 
in which weighted Klimontovich and related representations
are replaced by powerful technique from local nonlinear harmonic 
analysis, based on underlying symmetries of functional space such as affine or more general. 
The solution has the multiscale 
decomposition via nonlinear high-localized 
eigenmodes, which corresponds to the full multiresolution expansion in all 
underlying time/phase space scales. 
Starting from Vlasov-Poisson equations, we consider the approach based on 
multiscale variational-wavelet formulation. We give the explicit representation for all 
dynamical variables in the base of compactly supported wavelets/wavelet packets 
or nonlinear eigenmodes. 
Our solutions are parametrized by solutions of a number of reduced algebraical problems, one 
from which is nonlinear with the same degree of nonlinearity as initial problem and the others 
are the linear problems which correspond to the particular method of calculations inside 
concrete wavelet scheme. Because our approach started from variational formulation we can 
control evolution of instability on the pure algebraical level of reduced algebraical system of 
equations. We consider the following form of 
equations
\begin{eqnarray}
&&\Big\{\frac{\partial}{\partial s}+p_x\frac{\partial}{\partial x}+
             p_y\frac{\partial}{\partial y}-
\Big[k_x(s)x+\frac{\partial\psi}{\partial x}\Big]\frac{\partial}{\partial p_x}-
 \Big[k_y(s)y+\frac{\partial\psi}{\partial y}\Big]\frac{\partial}{\partial p_y}
  \Big\} f_b(x,y,p_x,p_y,s)=0, \\
&&\Big(\frac{\partial^2}{\partial x^2}+\frac{\partial^2}{\partial y^2}\Big)\psi=
-\frac{2\pi K_b}{N_b}\int \ud p_x \ud p_y f_b,\quad
\int\ud x\ud y\ud p_x\ud p_y f_b=N_b.\nonumber
\end{eqnarray} 		
The corresponding Hamiltonian for transverse single-particle motion is given by 
$ 
H(x,y,p_x,p_y,s)=\frac{1}{2}(p_x^2+p_y^2) 
                   +\frac{1}{2}[k_x(s)x^2 
+k_y(s)y^2]+
    H_1(x,y,p_x,p_y,s)+\psi(x,y,s), 
$		
where $H_1$  is nonlinear (polynomial/rational) part of the full Hamiltonian and corresponding 
characteristic equations are: 
$
\frac{\ud^2x}{\ud s^2}+k_x(s)x+\frac{\partial}{\partial x}\psi(x,y,s)=0,\qquad
\frac{\ud^2y}{\ud s^2}+k_y(s)y+\frac{\partial}{\partial y}\psi(x,y,s)=0.
$
We obtain our multiscale/multiresolution representations for solutions of these 
equations via variational-wavelet approach. We decompose the solutions as 
$
f_b(s,x,y,p_x,p_y)=\sum^\infty_{i=i_c}\oplus\delta^if(s,x,y,p_x,p_y),\quad
\psi(s,x,y)=\sum^\infty_{j=j_c}\oplus\delta^j\psi(s,x,y),\quad
x(s)=\sum^\infty_{k=k_c}\oplus\delta^kx(s),\quad
y(s)=\sum^\infty_{\ell=\ell_c}\oplus\delta^\ell y(s),
$
where set 
$
(i_c,j_c,k_c,\ell_c)
$
corresponds to the coarsest level of resolution $c$  in the full multiresolution 
decomposition
$
V_c\subset V_{c+1}\subset V_{c+2}\subset\dots
$
Introducing detail space $W_j$   as the orthonormal complement of  $V_j$ with respect to 
$
V_{j+1}: 
V_{j+1}=V_j\bigoplus W_j,
$
         we have for 
$
f, \psi, x, y \subset L^2({\bf R}):
$
$\quad
L^2({\bf R})=\overline{V_c\displaystyle\bigoplus^\infty_{j=c} W_j}.
$
In some sense it is some generalization of the old $\delta F$  approach [3]. Let $L$  be an 
arbitrary (non) linear differential/integral operator with matrix dimension $d$, which acts,
according to (1), on 
some set of functions
$
\Psi\equiv\Psi(s,x)=\Big(\Psi^1(s,x),\dots,\Psi^d(s,x)\Big),\ 
 s,x \in\Omega\subset{\bf R}^{n+1}
$
         from  $L^2(\Omega)$: 
$\quad
L\Psi\equiv L(R(s,x),s,x)\Psi(s,x)=0,
$
where  $x$ are the generalized space coordinates or phase space coordinates, and  $s$ is 
"time" coordinate. As a result the solution of equations (1) has 
the following full multiscale/multiresolution decomposition via nonlinear high-localized 
eigenmodes: 

\begin{eqnarray}\label{eq:z}
\Psi(s,{\bf x})&=&\sum_{(i,j)\in Z^2}a_{ij}{\bf U}^i\otimes V^j(s,{\bf x}),\quad {\bf x}=(x,y,p_x,p_y)\\
V^j(s)&=&V_N^{j,slow}(s)+\sum_{l\geq N}V^j_l(\omega_ls), \ \omega_l\sim 2^l, \quad
{\bf U}^i({\bf x})={\bf U}_M^{i,slow}({\bf x})+
\sum_{m\geq M}{\bf U}^i_m(k_m{\bf x}), \ k_m\sim 2^m, \nonumber
\end{eqnarray}

\begin{vchfigure}[htb]
\centering 
\includegraphics[width=65mm]{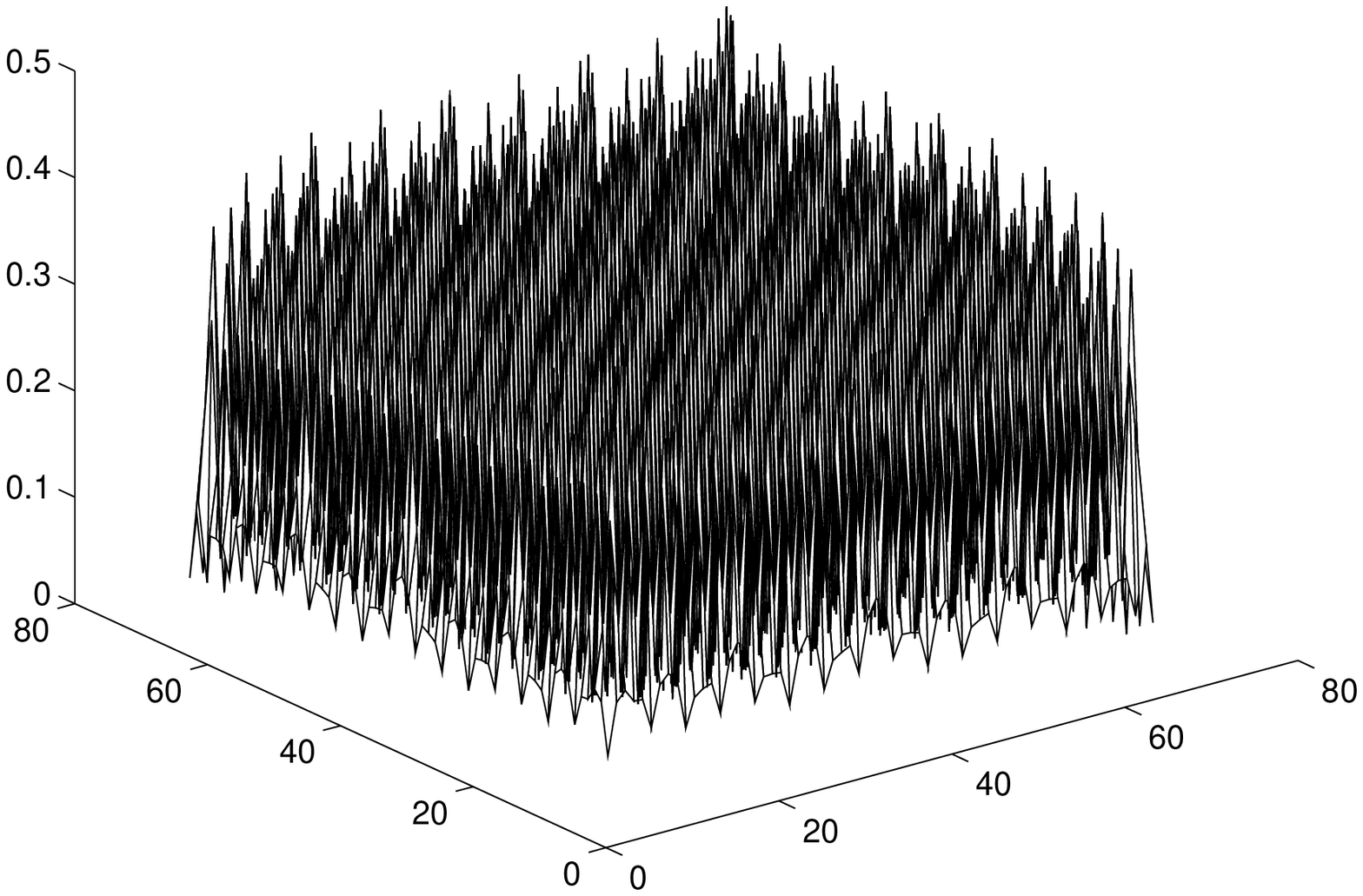} 
\vchcaption{Chaotic pattern.}
\end{vchfigure}

\begin{vchfigure}[htb]
\centering
\includegraphics[width=65mm]{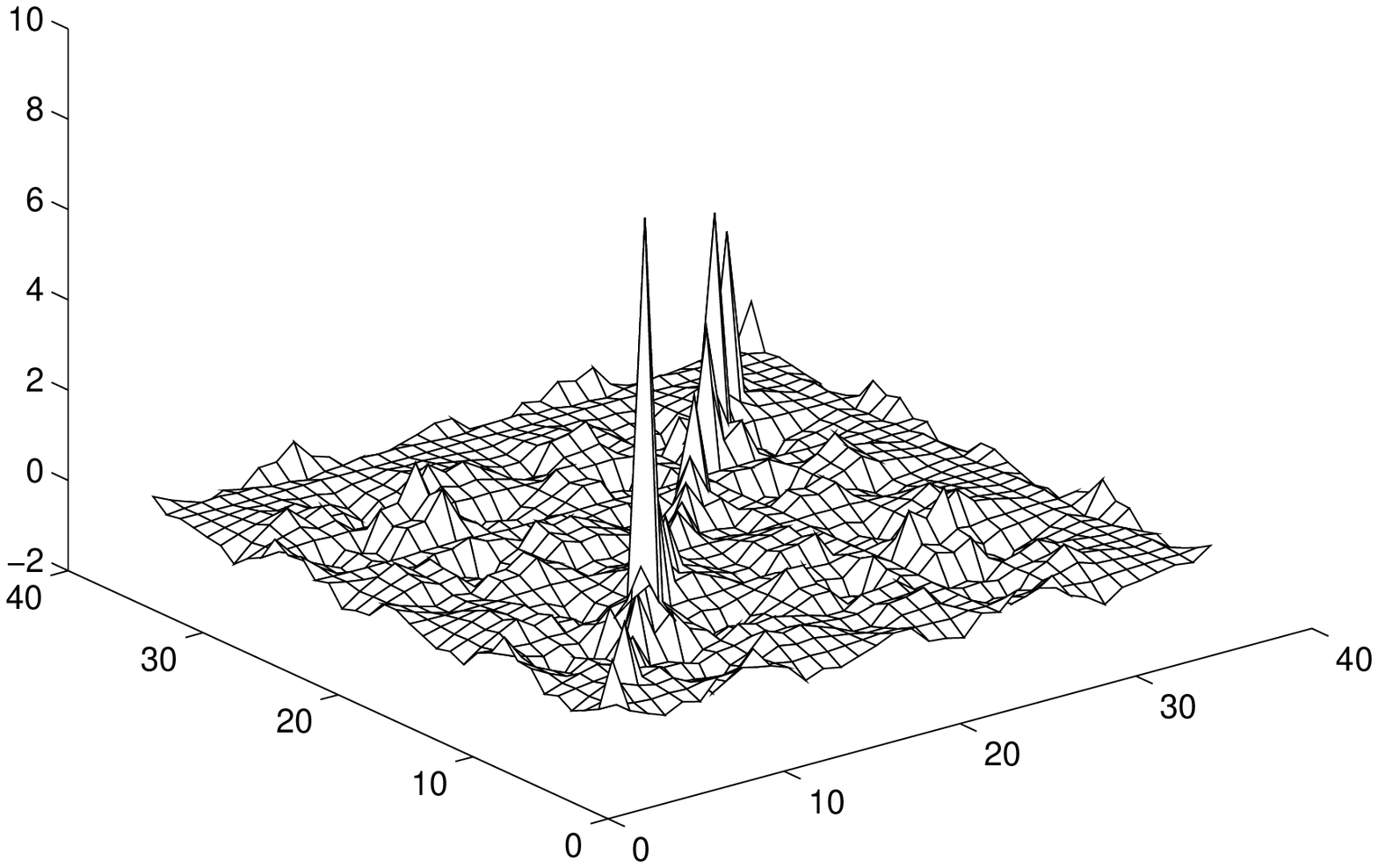} 
\vchcaption{Localized pattern(waveleton): energy confinement state.}
\end{vchfigure} 

It should be noted that such representations give the best possible 
localization properties in the corresponding (phase)space/\-ti\-me coordinates. In contrast with 
other approaches these formulas do not use perturbation technique or linearization procedures. 
Modeling demonstrates the appearance of stable patterns formation from high-
localized coherent structures or chaotic behaviour (Fig. 1). 
Such analysis and modeling describes, in principle, a 
scenario for the generation of controllable localized (meta) stable 
fusion-like state or waveleton (Fig. 2).
Definitely, chaotic-like unstable partitions/states (Fig. 1) dominate during non-equilibrium 
evolution. It means that (possible) localized (meta) stable partitions have measure equal to 
zero a.e. on the full space of hierarchy of partitions defined on a domain of the definition in 
the whole phase space. Nevertheless, our scheme 
give some chance to build the controllable 
localized state (Fig. 2) starting from initial chaotic-like 
partition via process of controllable self-organization.
Of course, such confinement states, waveleton, characterized by zero 
measure and minimum entropy, can be only metastable. But these long-living fluctuations can 
and must be very important from the practical point of view, because the averaged time of 
existence of such states may be even more than needed for practical realization, e.g., in 
controllable fusion processes.

\end{document}